\documentstyle [12pt]{article}
\def\fnote#1#2{\begingroup\def\thefootnote{#1}\footnote{#2}\addtocounter
{footnote}{-1}\endgroup}
\def\inbar{\vrule height1.5ex width.4pt depth0pt}
\def\IB{\relax{\rm I\kern-.18em B}}
\def\IC{\relax\,\hbox{$\inbar\kern-.3em{\rm C}$}}
\def\ID{\relax{\rm I\kern-.18em D}}
\def\IE{\relax{\rm I\kern-.18em E}}
\def\IF{\relax{\rm I\kern-.18em F}}
\def\IG{\relax\,\hbox{$\inbar\kern-.3em{\rm G}$}}
\def\IH{\relax{\rm I\kern-.18em H}}
\def\II{\relax{\rm I\kern-.18em I}}
\def\IK{\relax{\rm I\kern-.18em K}}
\def\IL{\relax{\rm I\kern-.18em L}}
\def\IM{\relax{\rm I\kern-.18em M}}
\def\IN{\relax{\rm I\kern-.18em N}}
\def\IO{\relax\,\hbox{$\inbar\kern-.3em{\rm O}$}}
\def\IP{\relax{\rm I\kern-.18em P}}
\def\IQ{\relax\,\hbox{$\inbar\kern-.3em{\rm Q}$}}
\def\IR{\relax{\rm I\kern-.18em R}}
\def\ZZ{\relax{\sf Z\kern-.4em Z}}
\def\fnote#1#2{\begingroup\def\thefootnote{#1}\footnote{#2}\addtocounter
{footnote}{-1}\endgroup}
\def\beq{\begin{equation}}
\def\eeq{\end{equation}}
\def\bea{\begin{eqnarray}}
\def\eea{\end{eqnarray}}
\def\lleq#1{\label{#1}\eeq}
\let\nn=\nonumber

\def\notin{\ \hbox{{$\in$}\kern-.51em\hbox{/}}}
\def\a{\alpha}        
     
     \def\si{\sigma}
   \def\th{\theta}

   \def\cM{{\cal M}}

   \def\bz{{\bar z}}

\def\bth{{\bar \theta}}

\def\picture #1 by #2 (#3){
  $$\vbox to #2{
    \hrule width #1 height 0pt depth 0pt
    \vfill
    \special{picture #3}}$$\vspace{-0.5cm}}
\def\scaledpicture #1 by #2 (#3 scaled #4){{
 \dimen0=#1 \dimen1=#2
 \divide\dimen0 by 1000 \multiply\dimen0 by #4
 \divide\dimen1 by 1000 \multiply\dimen1 by #4
 \picture \dimen0 by \dimen1 (#3 scaled #4)}}

\begin{document}

\hfill {
{HD--THEP--92--29}}
\vskip .05truein
\hfill {
{September 1992}}
\vskip 1truein
\centerline{\large  CRITICAL SUPERSTRING VACUA FROM NONCRITICAL MANIFOLDS}
\centerline{\sc A Novel Framework for String Compactification}

\vskip .9truein
\centerline{\sc Rolf Schimmrigk
     \fnote{\diamond}{Email address: q25@vm.urz.uni-heidelberg.de}}


\vskip .3truein
\centerline{\it Institut f\"ur Theoretische Physik,
                Universit\"at Heidelberg}
\centerline{\it Philosophenweg 16, 6900 Heidelberg, FRG}

\vskip 1.5truein
\centerline{\bf ABSTRACT}
\vskip .2truein

A new framework is found for the compactification of supersymmetric string
theory. It is shown that the massless spectra of Calabi--Yau manifolds of
complex dimension $D_{crit}$ can be derived from noncritical manifolds of
complex dimension $2k+D_{crit}$, $k\geq 0$. These higher dimensional
manifolds are spaces whose nonzero Ricci curvature is quantized in a
particular way. This  class is more general than that of Calabi--Yau
manifolds because it contains spaces that correspond to critical string
vacua with no K\"ahler deformations, i.e. no antigenerations, thus
providing mirrors of rigid Calabi--Yau manifolds.
The constructions introduced here lead to new insights into the relation
between exactly solvable models and their mean field theories on the one
hand and Calabi--Yau manifolds on the other.
They  also raise fundamental questions about the Kaluza--Klein concept of
string compactification, in particular regarding the r\^{o}le played by
the dimension of the internal theories.


\renewcommand\thepage{}
\vfill
\eject

\parskip .1truein
\baselineskip=17pt    
\parindent=20pt
\pagenumbering{arabic}

\noindent
{\sc 1. Introduction}

\noindent
It is believed that the heterotic string without torsion can propagate
consistently in a manifold only if this manifold is complex, K\"ahler and
admits a covariantly constant spinor, i.e. has vanishing first Chern class.
Manifolds of this type, so--called Calabi--Yau manifolds, are examples
of left--right symmetric string vacua with $N=2$ supersymmetry on the
world sheet. It is further believed that the configuration space of such
groundstates features an important symmetry, not at all manifest in the
construction of the superstring: mirror symmetry. The predictions of
this symmetry, which has been discovered in the context of
Landau--Ginzburg vacua in [1] and proven to exist in this framework in
ref. [2], have been shown to be correct in all computations performed
so far [3,4]. Independent evidence for this symmetry has been found in the
context of orbifolds of exactly solvable tensor models by Greene and
Plesser [5].

Mirror symmetry creates a puzzle.
There are well--known Calabi--Yau vacua which are rigid, i.e. they do
not have string modes corresponding to complex deformations of the
manifold, fields that describe generations in the low energy theory.
Since mirror symmetry exchanges complex deformations and K\"ahler
deformations of a manifold, the latter describing the antigenerations
seen by a four--dimensional observer, it would seem that the
mirror of a rigid Calabi--Yau manifold
cannot be K\"ahler and hence does not describe a consistent string vacuum.
It follows that the class of Calabi--Yau manifolds is not the
appropriate setting in which to discuss mirror symmetry and the
question arises  what the proper framework might be.

It is the purpose of this article to introduce a new class of manifolds
which generalizes the class of Calabi--Yau spaces of complex dimension
$D_{crit}$ in a natural way. The manifolds involved are of complex
dimension $(2k+D_{crit})$ and have a positive first Chern class which
is quantized in multiples of the degree of the manifold.
Thus they do not describe, a priori, consistent string groundstates.
Surprisingly, however, it is possible to derive from these higher dimensional
manifolds the spectrum of critical string vacua. This
can be done not only for the generations but also for the antigenerations.
For particular types of these new manifolds it is in fact possible to
construct $D_{crit}$--dimensional Calabi--Yau manifolds directly from the
$(2k+D_{crit})$--dimensional spaces.

This new  class of manifolds is, however, not in one to one
correspondence with the class of Calabi--Yau manifolds as it contains
 manifolds which describe string vacua that do not contain massless modes
corresponding to antigenerations.
It is precisely this new type of manifold that is
needed in order to construct mirrors of rigid Calabi--Yau manifolds
 without generations. The results presented in this article suggest
that the noncritical manifolds described here are no less physical
than critical manifolds and indeed define
the appropriate generalization of the Calabi--Yau framework
of string compactification. They also lead to important ramifications
regarding the relation between Landau--Ginzburg theories and
critical manifolds.

\vfill \eject

\noindent
{\sc 2. Higher Dimensional Manifolds with Quantized Positive First
Chern Class}

\noindent
Consider the class of manifolds of complex dimension $N$ embedded in a
weighted projective space $\IP_{(k_1,\dots, k_{N+2})}$ as hypersurfaces
\beq
\cM_{N,d}=~ \IP_{(k_1,k_2,\dots \dots ,k_{N+2})}[d] = ~
\{p(z_1,\dots,z_{N+2})=0\}~\cap ~\IP_{(k_1,\dots, k_{N+2})} \nn
\lleq{phyp}
defined as the zero set of some transverse polynomial $p$ of degree $d$.
Here the integers $k_i$ describe the weights of the ambient weighted
projective space. $\IP_{(k_1,k_2,\dots \dots ,k_{N+2})}[d]$ is called a
configuration.
Assume that for the hypersurfaces (\ref{phyp}) the weights $k_i$ and the
degree $d$ are related via the constraint
\beq
\sum_{i=1}^{N+2} k_i =  Qd,
\lleq{pquant}
where $Q$ is a positive integer. Relation (\ref{pquant}) is the defining
property of the class of manifolds I will consider in this article.
It is a rather
restrictive condition in that it excludes many types of varieties which
are transverse and even smooth but are not of physical relevance
\fnote{1}{The erudite reader will recognize that this definition is
          rather natural in the context of Landau--Ginzburg
         compactification with an arbitrary number of scaling fields
         as will become clear below. A particularly simple
      manifold in this class, the cubic sevenfold $\IP_8[3]$, has been
      the subject of recent investigations [6--8].}.

Alternatively, manifolds of the type above may be characterized
via a curvature constraint. Because of (\ref{pquant}) the first Chern
class is given by
\beq
c_1(\cM_{N,d}) =(Q-1)~d~h
\lleq{c1quant}
where $h$ is the pullback of the K\"ahler form of the ambient space. Hence
the first Chern class is quantized in multiples of the degree of the
hypersurface. For $Q=1$ the first Chern class vanishes and
the manifolds for which condition (\ref{pquant}) holds are
Calabi--Yau manifolds, defining
consistent ground states of the supersymmetric closed string.
For $Q > 1$ the first Chern class is nonvanishing and therefore these
manifolds cannot possibly describe vacua of the critical string, or so
it seems.

It will be shown below that these spaces are closely
related to string vacua of critical dimension
\beq
D_{crit} = N-2(Q-1)
\lleq{critdim}
i.e. the critical dimension is offset by twice the coefficient of the
first Chern class of the normal bundle.
The evidence for this is twofold. First it is possible to
derive from these higher dimensional manifolds the massless spectrum
of critical vacua. Furthermore it is shown that it is possible to construct
Calabi--Yau manifolds $M_{CY}$ of dimension $D_{crit}$  and complex
codimension $codim_{\IC} (M_{CY}) =Q$
directly from certain subclasses of hypersurfaces of type (\ref{pquant}).
In terms of the critical dimension and
the codimension the class of manifolds to be investigated below can be
described as the projective configurations
\beq
\IP_{(k_1,\dots ,k_{(D_{crit}+2Q)})}
\left[\frac{1}{Q}\sum_{i=1}^{D_{crit}+2Q} k_i\right] .
\lleq{newmfs}

The class defined by
(\ref{newmfs}) contains manifolds with no antigenerations.
Hence it is necessary to have some way other than Calabi--Yau manifolds
to represent string ground states in order to establish a relation
between such higher dimensional manifolds and string vacua. One possible
 way to achieve this is via
Landau--Ginzburg theories: manifolds of type (\ref{phyp}) can be
viewed as a projectivization via a weighted equivalence of an affine
noncompact hypersurface  defined by the same polynomial
\beq
\IC_{(k_1,...,k_{N+2})}\left[d\right] \ni \{p(z_1,...,z_{N+2})=0\}.
\lleq{affvar}
Because the polynomial $p$ is assumed to be transverse in the projective
ambient
space the affine variety has a very mild singularity: it has an isolated
singularity at the origin, defining what is called a catastrophe in the
mathematics literature.

The complex variables $z_i$ parametrizing the ambient space are to be
viewed as the field theoretic limit, $\varphi_i(z,\bz) = z_i$, of the
lowest components of the order parameters
$\Phi_i(z_i,\bz_i,\th^{\pm}_i,\bth^{\pm}_i)$
described by chiral $N=2$ superfields of a 2--dimensional
Landau--Ginzburg theory.
It was the important insight of Martinec [9] and Vafa and Warner [10]
that such Landau--Ginzburg theories are useful for the understanding of
 string
vacua and also that much information about such ground states is already
encoded in the associated catastrophe (\ref{affvar}). A crucial piece of
information about a vacuum, e.g., is its central charge. Using a result
from singularity theory, it is easy to derive that the central charge of
the conformal fixed point of the LG theory is
$c=3\sum_{i=1}^{N+2} \left(1-2q_i\right)$,
where $q_i =k_i/d$ are the U(1) charges of the superfields. It is
furthermore  possible
to derive the massless spectrum of the GSO projected fixed point of the
Landau--Ginzburg theory defining the string vacuum directly from the
catastrophe (\ref{affvar}) via a procedure described by Vafa [11].
The manifolds (\ref{newmfs}) therefore correspond to (in general
somewhat unconventional) Landau--Ginzburg theories with central
charge
\beq
c=3(N-2(Q-1))=3D_{crit}
\eeq
where the relation (\ref{critdim}) has been used.

In certain benign situations the subring of monomials
of charge 1 in the chiral ring describes the generations of the
vacuum [12]. For this to hold at all it is important that the GSO
projection is the canonical one with respect to the cyclic group $\ZZ_d$,
the  order of which is the degree $d$ of the superpotential
\fnote{2}{It does not hold for projections that involve orbifolds with
       respect to different groups such as those discussed in [13]. This is
       to be expected as these modified projections
       can be understood as orbifolds of canonically constructed
       vacua. The additional moddings generate singularities the resolution
       of  which introduces, in general,
       additonal modes in both sectors, generations and antigenerations.}.
Thus the generations are easily derived for this subclass of
theories in (\ref{newmfs})  because the polynomial ring is identical
to the chiral ring of the corresponding Landau--Ginzburg theory.
In general a more sophisticated analysis, involving the resolution of
higher dimensional singularities, will have to be done [14].

It remains to extract the second cohomology. In a Calabi--Yau manifold
there are no holomorphic  1--forms and hence all of the second cohomology
is in $H^{(1,1)}$. Because of Kodaira's vanishing theorem the same is
true for manifolds with positive  first Chern class and therefore for
the manifolds under discussion.
At first sight it might appear hopeless
to find a construction which would allow one to relate
the antigenerations of the critical vacuum to the (1,1)--cohomology
of the higher dimensional manifold because of the following example.
Consider the
orbifold $T_1^3/\ZZ_3^2$ where the two actions are defined as
$(z_1,z_4) \longrightarrow  (\a z_1,\a^2 z_4)$, all other coordinates
invariant
and $(z_1,z_7) \longrightarrow (\a z_1,\a^2 z_7) $, all other invariant.
Here $\a$ is the third root of unity.
The resolution of the singular orbifold leads to a Calabi--Yau manifold with
 84 antigenerations and no generations [15].
This is precisely the mirror flipped spectrum of the exactly solvable tensor
model $1^9$ of 9 copies of $N=2$ superconformal minimal models at level
$k=1$ [16] which can be described in terms of the Landau--Ginzburg
potential $W=\sum \Phi_i^3$ which belongs to the configuration
$\IC_{(1,1,1,1,1,1,1,1,1)}[3]$.

This Landau--Ginzburg theory clearly is a mirror candidate for the
resolved torus orbifold just mentioned [6-8]
\fnote{3}{A detailed comparison of the Yukawa couplings of the
         Landau--Ginzburg theory with those of the `instanton corrected'
       resolved orbifold has been performed in [7].}
and the  question arises whether a manifold corresponding to
this LG potential can be found. Since the theory does not contain modes
corresponding to (1,1)--forms it seems that the manifold cannot be
K\"ahler and hence not projective. Thus it appears that the 7--dimensional
 manifold $\IP_8[3]$ whose polynomial ring is  identical to the chiral ring
of the LG theory is merely useful as an auxiliary device
in order to describe one sector of the critical LG string vacuum:
Even though there exists a precise identity between the Hodge numbers in the
middle cohomology group of the higher dimensional manifold and the middle
dimensional cohomology of the Calabi--Yau manifold this is not
the case for the second cohomology group.

\vskip .3truein
\noindent
{\sc 3. Relation between Critical and Noncritical Manifolds}

\noindent
It turns out that by looking at the manifolds of the type
described by (\ref{newmfs})  in a particular way it is indeed possible
to extract the second cohomology in a canonical manner (even if there is
{\it none}).
The way this works is as follows: the manifolds
(\ref{newmfs}) will, in general, not be described by smooth spaces but
will have singularities which arise from the projective identification.
The basic idea now is to associate the existence of antigenerations in
a {\it critical}
string vacuum with the existence of singularities in these higher
dimensional {\it noncritical} spaces.

Since the structure of these geometrical singularities depends on the
precise form of the polynomial constraint it is difficult to
prove the correctness of this idea in full generality.
Instead I will, in the following, make the ideas involved more precise and
illustrate  how they work with a few particularly simple classes of
theories, leaving a more detailed investigation of other types of
manifolds to a more extensive discussion [14]. As an
unexpected bonus this derivation will provide new insight
into the Landau--Ginzburg/Calabi--Yau connection.

It is useful to first consider an example in some detail. The GSO
projected LG theory based on the superpotential
\beq
W= \sum_{i=1}^3 \left( \Phi_i^3\Psi_i + \Psi_i^3\right) + \Psi_4^3
\eeq
describes a vacuum with 35 generations and 8 antigenerations.
Associated to this groundstate is the affine configuration
$\IC_{(2,3,2,3,2,3,3)}[9]$
which induces, via projectivization, a 5--dimensional weighted
hypersurface  $\IP_{(2,2,2,3,3,3,3)}[9]$. This compact manifold has
two types of orbifold singularities:
\bea
\ZZ_3 &:& \IP_3[3] \ni \{p_1=\sum_{i=1}^4 x_i^3=0\} \nn \\
\ZZ_2 &:& \IP_2.
\eea
The $\ZZ_3$--singular set is a smooth cubic surface which supports
 seven (1,1)--forms  whereas the
$\ZZ_2$--singular set is just the projective plane and therefore
adds one further (1,1)--form.  Hence the singularities induced on the
hypersurface by the singularities of the ambient weighted projective space
give rise to a total of eight  (1,1)--forms. A simple count leads to the
result that the subring of monomials of charge 1 is of dimension 35.
Thus we have derived the
spectrum of the critical theory from the noncritical manifold
$\IP_{(2,2,2,3,3,3,3)}[9]$.

It is presumably possible to derive this result via a surgery process
on the singular space, but more important is, at this point,
 that the idea introduced above of relating the spectrum of the string
vacuum to the singularity structure of the noncritical manifold also
makes it possible to derive from these higher dimensional
manifolds the Calabi--Yau manifold of critical dimension! This leads to
a canonical prescription which allows to pass from the
Landau--Ginzburg theory to its geometrical counterpart when the model
has antigenerations.

This works as follows: Recall that the structure of the singularities of
the weighted hypersurface
just involves part of the superpotential, namely the cubic polynomial
$p_1$ which determined the $\ZZ_3$ singular set described by a surface.
The superpotential thus splits naturally into the two parts
$p=p_1 + p_2$, where $p_2$ is the remaining part of the polynomial. The
idea is to consider the product $\IP_3[3] \times \IP_2$, where the
factors are determined by
the singular sets of the higher dimensional space and to impose on this
4--dimensional space a constraint described by  the remaining part of the
polynomial which did not
take part in constraining the singularities of the ambient space. In the
case at hand this leaves a polynomial of bidegree $(3,1)$ and hence we
are lead to a manifold embedded in
\beq
\matrix{\IP_2 \hfill \cr \IP_3\cr}
\left[\matrix{3&0\cr 1&3\cr}\right] \ni
\left\{ \begin{array}{c l}
         p_1~= &y_1^3x_1 + y_2^3x_2 +y_3^3x_3 =0\\
         p_2~= &\sum_{i=1}^4 x_i^3 =0
        \end{array}
\right\}.
\lleq{ex3mine}
But this is a well known Calabi--Yau manifold of complex dimension 3, first
constructed in [17]!

The ideas just described are general.
A subclass of manifolds of a different type which can be discussed in
this framework rather naturally is defined by the projective configurations
\beq
\IP_{(2k,K-k,2k,K-k,2k_3,2k_4,2k_5)}[2K]
\lleq{niceclass}
where $K=k+k_3+k_4+k_5$ and it is assumed, for simplicity,
 that  $K/k$ and $K/k_i$ are integers. The  potentials are
\beq
W=\sum_{i=1}^2(x_i^{K/k}+x_iy_i^2) +x_3^{K/k_3} +x_4^{K/k_4} +x_5^{K/k_5}
\eeq
and the singularities in these manifolds are of two types,
\bea
\ZZ_2 &:&~~\IP_{(k,k,k_3,k_4,k_5)}[K] \ni
 \left\{ p_1 = \sum_{i=1}^5 x^{K/k_i} =0 \right\} \nn \\
\ZZ_{K-k} &:&~~ \IP_1.
\eea
The $\ZZ_2$--singular set is 3--fold with positive first Chern class
embedded in weighted $\IP_4$ whereas the $\ZZ_{K-k}$--singular set is
just the sphere $S^2 \sim \IP_1$.

In complete analogy with the previous discussion the manifolds in this
class lead to critical manifolds embedded in
\beq
\matrix{\IP_1 \hfill \cr \IP_{(k,k,k_3,k_4,k_5)}\cr}
\left[\matrix{2&0\cr k&K\cr}\right] \ni
\left\{ \begin{array}{c l}
       p_1= & y_1^2x_1 + y_2^2x_2 =0  \\
       p_2= & x_1^{K/k}+ x_2^{K/k} + x_3^{K/k_3} + x_4^{K/k_4} + x_5^{K/k_5}=0
        \end{array}
\right\}.
\lleq{s3class}
That this correspondence is in fact  correct  can be inferred from the
work of [18]
where it was shown that these codimension--2 weighted CICYs correspond
to $N=2$ minimal exactly solvable tensor models  of the type
\beq
\left[ 2\left(\frac{K}{k}-1\right)\right]_D^2 \cdot
\prod_{i=3}^5  \left(\frac{K}{k_i}-2\right)_{A}.
\eeq
where the subscripts indicate the affine invariants chosen for the
individual levels.

The general picture that emerges from these constructions then is the
following: Embedded in the higher dimensional manifold is a submanifold
which is fibered, the base and the fibers being determined by the
singular sets of the ambient manifold. The Calabi--Yau manifold itself
is a hypersurface embedded in this fibered  submanifold. A heuristic
sketch of the geometry is shown in the Figure 1.



The examples above illustrate the simplest situation that can appear.
In more complicated manifolds the singularity structure will consist
of hypersurfaces whose fibers and/or base themself are fibered, leading to an
iterative procedure. The submanifold to be considered will, in those cases,
be of codimension larger than one and the Calabi--Yau
manifold will be described by a submanifold with codimension
larger than one as well. In the most general situation the fiber bundle will
presumably not be simply
a product bundle as in the previous examples but will presumably involve
nontrivial twists.

The relation between the noncritical manifolds of type (\ref{newmfs}) and
critical string vacua is not 1--1. Indeed, by using the construction of
`splitting' and `contracting' Calabi--Yau manifolds introduced in [19] it
follows that noncritical manifolds of different dimensions can lead to one
and the same critical vacuum. Thus there exist nontrivial relations between
the spaces of type (\ref{newmfs}). A more detailed discussion of these
aspects will appear in [14].

In the framework described above it becomes clear what is special about
string vacua that do not have modes corresponding to antigenerations.
Consider again the example related to the tensor
model $1^9$. Its LG theory describes an affine cubic surface in $\IC_9$ the
naive compactification of which leads to
\beq
\IP_8[3]\ni \{p(z_1,...,z_9)=\sum_{i=1}^9 z_i^3=0\}.   \nn
\lleq{ex1}
Counting monomials leads to the spectrum of 84 generations found previously
for the corresponding string vacuum and because this manifold is
{\it smooth} {\it no} antigenerations are expected in this model!
Hence there does not exist a Calabi--Yau manifold that describes this
ground state.
A second theory in the space of all LG vacua with no antigenerations is
\beq
(2^6)_{A^6}^{(0,90)} \sim \IC_{(1,1,1,1,1,1,2)}[4] \ni
\left\{\sum_{i=1}^6 z_i^4 + z_7^2 =0\right\}                      \nn
\lleq{ex2}
with an obviously smooth manifold $\IP_{(1,1,1,1,1,1,2)}[4] $.

\vskip .3truein
\noindent
{\sc 4. Generalization to Arbitrary Critical Dimensions}

\noindent
Even though the examples discussed in the previous section are all
concerned with critical vacua of central charge $c=9$ and the way they are
related to the new class of noncritical spaces of dimension $3+2k$, it should
be clear that these considerations are not specific to
this particular set of string groundstates. Instead of considering
`compactifications' of the heterotic string down to the
physical dimension, 4, we can contemplate compactifying down to 2, 6 or
8 dimensions, or else, discuss the class of manifolds type
(\ref{newmfs}) independently from string considerations.

To illustrate this point consider the infinite class of
$(n+1)$--dimensional manifolds
\beq
\IP_{(2,n-1,2,n-1,2,\dots ,2)}[2n]
{}~~\ni ~~\{p=\sum_{i=1}^2(x_i^n+x_iy_i^2) +x_3^n +\cdots +x_{n+1}^n=0\}.
\eeq
According to the ideas of the previous sections these spaces are
related to Calabi--Yau manifolds embedded in products of projective spaces
\beq
\matrix{\IP_1\cr \IP_n\cr}\left[\matrix{2&0\cr 1&n\cr}\right] \ni
\left\{ \begin{array}{c l}
   p_1 =& y_1^2x_1+y_2^2x_2=0 \\
p_2 =& \sum_{i=1}^{n+1} x_i^n =0
        \end{array}
\right\},
\eeq
corresponding to critical vacua with central charge $c=3(n-1)$. Particularly
interesting is the case of K3 because it shows that the procedure also works
for smooth configurations involving non--Fermat type potentials. It should be
emphasized that the constructions of section 3 are not restricted to the
classes of spaces to which I have confined the present analysis for the sake
of brevity. A more complete discussion is involved and will appear
elsewhere [14].

\vskip .3truein
\noindent
{\sc 5. Conclusion}

\noindent
Mirror symmetry cannot be understood in the framework of Calabi--Yau manifolds.
Assuming that mirror symmetry is indeed a symmetry of the space of left--right
symmetric vacua and that the geometrical framework is general enough would lead
one to suspect the existence of a space of a new
type of noncritical manifolds which contain information about critical vacua,
such as the mirrors of rigid Calabi--Yau manifolds. Mirrors of spaces with
both sectors, antigenerations and generations, are, however, again of
Calabi--Yau type
and hence the noncritical manifolds which correspond to such groundstates
should make contact with Calabi--Yau manifolds in some manner.

It has been  shown that the class of higher dimensional K\"ahler
manifolds of type (\ref{newmfs}) with positive first Chern class,
quantized in a particular way,
generalizes the framework of Calabi--Yau vacua in the desired way: For
particular types of such noncritical manifolds Calabi--Yau manifolds of
critical dimension are embedded algebraically in a fibered submanifold.
For string vacua which cannot be described by
K\"ahler manifolds and which are mirror candidates of rigid Calabi--Yau
manifolds the higher dimensional manifolds still lead to the
spectrum of the critical vacuum and a rationale emerges that explains why a
Calabi--Yau representation is not possible in such theories.
Thus these manifolds of dimension $c/3 +2k$
 define an appropriate framework in which to discuss mirror symmetry.

There are a number of important consequences that follow from the results
of the previous sections. First it should be realized that the relevance
of noncritical manifolds suggests the generalization of a conjecture regarding
the relation between superconformal field theories with N=1 spacetime
supersymmetry and central charge $c=3D$, where $D\in \IN$,  on the one hand
and K\"ahler manifolds
of complex dimension $D$ with vanishing first Chern class on the other.
It was suggested by Gepner [20] that this relation is 1--1.
It follows from the results above that
instead superconformal theories of the above type are in correspondence
with K\"ahler manifolds of dimension $c/3 +2k$ with a first Chern
class quantized in multiples of the degree.

A second consequence is that the ideas of section 3 lead, for a large
class of Landau-Ginzburg theories, to a new canonical prescription for the
construction of the critical manifold, if it exists, directly from the 2D
field theory.

Recently Batyrev [21] introduced a new construction of mirrors of
Calabi--Yau manifolds based on dual polyhedra. His method appears to
apply only to manifolds defined by one polynomial in a weighted projective
space or products thereof. The method of toric geometry that is used in [21]
is however not restricted to Calabi--Yau manifolds and therefore the
constructions described in sections 3 and 4 lead to the exciting possibility
of extending Batyrev's results to Calabi--Yau manifolds of codimension larger
than one by proceeding via noncritical manifolds.

A final remark is that in this framework the role played by the dimension
of the manifolds  becomes of secondary importance. This is as it should be,
at least for an effective theory, which tests only matter content and
couplings. It is, however, somewhat mysterious that via ineffective
splittings [19], manifolds of different
dimension describe one and the same critical vacuum.

It is clear that the emergence in string theory of manifolds with
quantized first Chern class should be understood better. The results
presented here are a first step in this direction. They indicate
that these manifolds are not just  auxiliary devices but may be as
physical as  Calabi--Yau manifolds of critical dimension.
In order to probe the structure of these models in more depth it is
important to get further insight into the complete spectrum of
these theories and to compute the Yukawa couplings of the fields.
The spectra of the
higher dimensional manifolds contain additional modes beyond those that
are related to the generations and antigenerations of the critical vacuum
and the question arises what physical interpretation these fields afford.

A better grasp on the complete spectrum of these spaces should also
give insight into a different, if not completely independent, approach
toward a deeper understanding of these higher dimensional manifold, which
is to attempt the construction of consistent $\si$--models defined via
these spaces. Control of the spectrum will shed light
on the precise relation between the $\si$--models based on Calabi--Yau
manifolds and noncritical $\si$--models.

\vskip .3truein
\noindent
{\sc Acknowledgement}

\noindent
I'm grateful to CERN for hospitality and the theoretical theorists there
for lively discussions, in particular Per Berglund, Philip Candelas,
Wolfgang Lerche and Jan Louis. It is a pleasure to thank Herbert Clemens,
Tristan H\"ubsch, Cumrun Vafa and Nick Warner for conversations and
R.Hartshorne for asking the right question. I'm also grateful to the
Aspen Center of Physics for hospitality.

\vfill \eject

\noindent
{\sc References}
\begin{enumerate}
\item P. Candelas, M.Lynker and R.Schimmrigk,
      Nucl.Phys. {\bf B341}(1990)383
\item M.Lynker and R.Schimmrigk, Phys.Lett. {\bf 249B}(1990)237
\item P.Candelas, X.de la Ossa, P.Green and L.Parkes,
      Phys.Lett {\bf 258B}(1991)118; Nucl.Phys. {\bf B359}(1991)21
\item S.Katz, private communication
\item B.R.Greene and R.Plesser, Nucl.Phys. {\bf B338}(1990)15
\item P.Candelas, Talk at the Workshop on Geometry and Quantum
      Field Theory, Baltimore, March 1992
\item P.Candelas, E.Derrick and L.Parkes, in preparation
\item C.Vafa, {\it Topological Mirrors and Quantum Rings},
      Harvard preprint HUTP--91/A059
\item E.Martinec, Phys.Lett. {\bf 217B}(1989)431
\item C.Vafa and N.Warner, Phys.Lett. {\bf 218B}(1989)51
\item C.Vafa, Mod.Phys.Lett. {\bf A4}(1989)1169
\item P.Candelas, Nucl.Phys. {\bf B298}(1988)458
\item P.Berglund, B.R.Greene and T.H\"ubsch,
      {\it Classical vs. Landau--Ginzburg Geometry of Compactification},
      CERN preprint, CERN--TH--6381/92
\item R.Schimmrigk, work in progress
\item B.R.Greene, C.Vafa and N.Warner, Nucl.Phys. {\bf B324}(1989)371
\item D.Gepner, Nucl.Phys. {\bf B296}(1988)757
\item R.Schimmrigk, Phys.Lett. {\bf 193B}(1987)175
\item R.Schimmrigk, Phys.Lett. {\bf 229B}(1989)227
\item P.Candelas, A.Dale, C.A.L\"utken and R.Schimmrigk,
      Nucl.Phys. {\bf B298}(1988)493
\item D.Gepner, Phys.Lett. {\bf 199B}(1987)380
\item V.V.Batyrev, {\it Dual Polyhedra and Mirror Symmetry for Calabi--Yau
            Hypersurfaces in Toric Varieties}, University of Essen preprint
\end{enumerate}

\end{document}